\documentclass[aps,prl,twocolumn]{revtex4}
\usepackage{graphicx}
\usepackage{dcolumn}
\usepackage{amsmath}
\usepackage{amssymb}
\usepackage{subfigure,amsmath,verbatim,moreverb,bm}
\def\be{\begin{equation}}
\def\ee{\end{equation}}
\def\ber{\begin{eqnarray}}
\def\eer{\end{eqnarray}}

\def\rv{{\bf r}}

\def\fv{{\bf f}}

\def\dens{n}

\begin{document}
\title{Density functional theory for strongly-correlated bosonic and fermionic 
ultracold dipolar and ionic gases}
\author{F. Malet$^1$, A. Mirtschink$^1$, C. B. Mendl$^2$, J. Bjerlin$^3$, E. \"O. Karabulut$^{3,4}$, S. M. Reimann$^3$ and Paola Gori-Giorgi$^1$}
\affiliation{$^1$Department of Theoretical Chemistry and Amsterdam Center for 
Multiscale Modeling, FEW, Vrije Universiteit Amsterdam, The Netherlands\\
$^2$Mathematics Department, Technische Universit\"at M\"unchen, Germany\\
$^3$Mathematical Physics, Lund University, Sweden\\
$^4$Department of Physics, Faculty of Science, Selcuk University, Turkey}

\date{\today}

\begin{abstract}
We introduce a density functional formalism to study the ground-state properties of strongly-correlated dipolar and 
ionic ultracold bosonic and fermionic gases, based on the self-consistent combination of the weak and the strong coupling 
limits. Contrary to conventional density functional approaches, our formalism does not require a previous calculation of the 
interacting homogeneous gas, and it is thus very suitable to treat systems with tunable long-range interactions. Due to its asymptotic 
exactness in the regime of strong correlation, the formalism works for systems in which standard mean-field theories fail.  
\end{abstract}
\maketitle


\noindent{\it Introduction -- }In contrast with its widespread use and success in areas as 
diverse as quantum chemistry \cite{CohMorYan-CR-12}, materials science \cite{NeuTil-13} or 
semiconductor nanostructures \cite{ReiMan-RMP-02}, 
Density Functional Theory (DFT) has received relatively little attention 
in the very active field of ultracold atomic gases. It is well known that the 
Hohenberg-Kohn theorems, originally formulated in terms of the electron 
gas \cite{HohKoh-PR-64,KohSha-PR-65}, hold for both fermionic and bosonic systems,
as well as for interactions different than the Coulomb one. However, the lack of 
adequate density functionals has hindered the role of DFT in the study of ultracold 
atomic gases in favour of other well-established approaches, such as the widely 
used Gross-Pitaevskii (GP) method in the case of Bose gases. The latter is a mean-field 
approach and does not allow treating the effect of correlations, which play a crucial role
in many different phenomena occurring in ultracold quantum gases \cite{BloDalZwe-RMP-08}.
One then often turns to configuration-interaction (CI), quantum Monte Carlo (QMC) or 
Green's-function methods (for recent reviews, see, e.g., 
Refs.~\onlinecite{MinSucTosTosVig-PR-04,GioPitStri-RMP-08,BloDalZwe-RMP-08}).

The advantages of DFT are very well known from electronic-structure calculations \cite{CohMorYan-CR-12,NeuTil-13,ReiMan-RMP-02}. 
Being an in principle exact theory (although in practice relying on approximations), 
DFT allows to go beyond the mean-field description by taking into account correlations 
between the interacting particles. The Kohn-Sham (KS) mapping of the many-body problem into a 
non-interacting one makes it possible to apply the method to particle numbers orders or 
magnitude larger than those accessible with wave-function 
methods \cite{ReiMan-RMP-02,RonCavBelGol-JCP-06,WalWesLin-PRB-13,GhoGucUmrUllBar-NP-06,GhoGucUmrUllBar-PRB-07},
as well as to capture a large part of the effects of quantum statistics on the kinetic energy.
Initial efforts have already been made to generalize the formalism to bosonic and fermionic ultracold  
quantum gases \cite{Nun-JPB-99,KimZub-PRA-03,Bra-JPB-04,HaoChe-PRA-09,PingPilTroDai-Nat-12}. However,
the biggest challenge in Kohn-Sham DFT --for both electronic and ultracold atomic systems-- is the 
construction of good approximations for the so-called exchange-correlation functional \cite{CohMorYan-CR-12}, 
the term in the total energy describing the many-body effects beyond the Hartree level. The simplest approximations are those based on homogeneous 
interacting models, where analytical expressions for the exchange-correlation energy per particle are 
often available by fitting QMC \cite{CasSorSen-PRB-06,PingPilTroDai-Nat-12} or Bethe-Ansatz 
calculations \cite{LimSilOliCap-PRL-03,HaoChe-PRA-09,WanHaoZha-PRA-12}. Such so-called local-(spin-)density 
approximations (L(S)DA) have been generalized and applied to the study of 
Bose \cite{Nun-JPB-99,KimZub-PRA-03,Bra-JPB-04,WanZha-PRA-13} and 
Fermi \cite{HaoChe-PRA-09,PingPilTroDai-Nat-12,WanHaoZha-PRA-12} ultracold gases with short-range interactions 
in different geometries and dimensionalities with promising results.
The downside of L(S)DA-based approximations, however, is that they become unreliable for systems in 
which the interactions between the particles largely dominate over the kinetic energy and characteristic 
strong-correlation phenomena arise. Moreover, L(S)DA approaches may become unpractical for systems with 
tunable interactions, since for each different interaction a previous many-body calculation of the uniform 
system with that same interaction is needed, which can be demanding even in the simplest one-dimensional
case \cite{CasSorSen-PRB-06}. The tunability of the interactions is relevant, for example, in the interesting
case of ultracold dipolar quantum gases \cite{Bar-RPP-08,LahMenSanLewPfa-RPP-09,BarDalPupZol-CR-12}. These
systems have recently received considerable attention, particularly for low dimensionalities, where the collisional 
instability towards head-to-tail alignment of the dipoles can be suppressed (see the 
reviews \cite{Bar-RPP-08,LahMenSanLewPfa-RPP-09,BarDalPupZol-CR-12}). Remarkably, dipolar interactions also 
allow the control of reaction rates in ultracold chemistry (see Refs. \cite{Ni-Sci-08,Osp-Sci-10}). 
One-dimensional systems have been of interest in particular in connection with the discussion of crystalline 
structures in the strong coupling regime (see, for example, Refs. 
\onlinecite{AstMorChiBor-PRA-08,ZinWunMekHuaWanDem-PRA-11,DeuCreRei-PRA-10,DalPupZol-PRL-10,CitOriPalChi-PRA-07,WunZinMekHuaWanDem-PRL-11}). 

In this Letter, we introduce an alternative approximate functional to study ultracold 
gases with long-ranged interactions, based on the exact formulation of the 
strong coupling limit of the Hohenberg-Kohn density 
functional \cite{GorSeiVig-PRL-09,GorSei-PCCP-10}. This provides an effective single-particle potential in a rigorous 
and physical way \cite{MalGor-PRL-12,MalMirCreReiGor-PRB-13,MenMalGor-PRB-14}, 
without relying on calculations of the uniform system energy by means of other many-body
approaches. The formalism becomes asymptotically exact in the limits of both vanishing and extremely strong 
coupling. The latter is obviously the most interesting, because of the
plethora of important phenomena that are out of reach of mean-field theories and problematic for wave-function methods. 
Furthermore, our construction can be equally applied to fermionic and bosonic gases, by simply changing the 
kinetic-energy functional. 

\vspace{0.3cm}
\noindent{\it Formalism -- } The key idea of Kohn-Sham Density Functional Theory (KS DFT) is that the ground-state particle density $\dens(\rv)$ 
and energy $E_0$ of a general $N-$body system with interparticle interaction $v_{\rm  int}(\rv-\rv')$ and external confining potential 
$v_{\rm ext}(\rv)$ can be mapped (with some mathematical subtle caveats, see, e.g.~\cite{Lie-IJQC-83}) into a non-interacting problem with 
the same particle density $\dens(\rv)$, moving in an effective potential $v_{\rm KS}([\dens];\rv)=v_{\rm ext}(\rv)+v_{\rm Hxc}([\dens];\rv)$. 
The Hartree-exchange-correlation (Hxc) potential $v_{\rm Hxc}([\dens];\rv)$, which is a Lagrange multiplier for the density constraint, is 
obtained from the functional derivative $\frac{\delta E_{\rm Hxc}[\dens]}{\delta\dens(\rv)}$ of the difference in energy $E_{\rm Hxc}[\dens]$ 
between the interacting and the non-interacting systems.  Physically,  $v_{\rm Hxc}([\dens];\rv)$ transforms the many-body interaction 
effects on the density into a single-particle potential. The non-interacting system is usually chosen in order to capture the relevant 
effects of statistics on the kinetic energy: for both fermions and bosons it is defined as the system with density $\dens(\rv)$ and 
minimum possible kinetic energy with fermionic or bosonic statistics, respectively. The problem is then reduced to the self-consistent 
solution of the Kohn-Sham equations $[-\frac{1}{2}\nabla^2 + v_{\rm ext}(\rv)+v_{\rm Hxc}([\dens];\rv)]\phi_i(\rv) =
\varepsilon_i\phi_i(\rv)$, where the Hxc potential 
$v_{\rm Hxc}([\dens];\rv) =v_{\rm mf}([\dens];\rv) + v_{\rm
  xc}([\dens];\rv)$ is the sum of the Hartree mean-field (mf) and exchange-correlation (xc) 
contributions, the latter needing to be approximated. 
The KS single-particle orbitals $\phi_i$ determine the ground-state density of the system 
via the relation \cite{KohSha-PR-65} $\dens({\bf r}) = \sum_i n_i|\phi_i({\bf r})|^2$, where
$n_i$ is the occupancy of the $i$th orbital. For bosonic systems at zero temperature one has 
$n_0=N$, and by neglecting the exchange-correlation term in the KS potential one recovers 
the Gross-Pitaevskii equation, widely used for the study of dilute ultracold Bose gases with short-range interaction, 
where the effects of many-body correlations do not play an important role.

Here we introduce the ``strictly-correlated particles'' (SCP) functional $V_{\rm int}^{\rm SCP}[\dens]$, which is complementary to the KS 
non-interacting kinetic energy: it is defined as the minimum possible expectation value of the particle-particle interaction in a given 
density $\dens(\rv)$. For the Coulomb interaction, this functional has been widely studied \cite{Sei-PRA-99,SeiGorSav-PRA-07,GorSeiVig-PRL-09}, 
and it has been shown to be able to capture the physics of the strongly-correlated regime in model quantum wires and quantum 
dots \cite{MalGor-PRL-12,MalMirCreReiGor-PRB-13,MenMalGor-PRB-14}, yielding results beyond the mean-field level. The construction 
of $V_{\rm int}^{\rm SCP}[\dens]$ for a given density $\dens(\rv)$ is equivalent to an optimal transport (or mass transportation theory, a 
well-established field of mathematics and economics) problem with cost given by the interaction \cite{ButDepGor-PRA-12,CotFriKlu-CPAM-13}. 
While several rigorous results have appeared recently in the mathematics literature \cite{CodFriPas-INC-14,FriMenPasCotKlu-JCP-13,ColDepDiM-CJM-14,Pas-NL-13,Pas-JFA-13,GhoMoa-GFA-14,KimPas-SJMA-14,ColDiM-INC-13}, here we provide a simplified physical overview. The idea is that if we minimize the expectation of the interparticle interaction in a 
given density $\dens(\rv)$, we must have a non-zero probability to find one particle wherever $\dens(\rv)\neq 0$. The many-particle 
state is then a continuum superposition of ``strictly-correlated'' configurations $(\rv_1={\bf r},\rv_2=\fv_2(\rv),...,\rv_N=\fv_N(\rv))$, 
with $\rv$ spanning the whole region where $\dens(\rv)\neq 0$,
\begin{align}
& |\Psi_{\rm SCP}(\rv_1,\dots,\rv_N)|^2 = \frac{1}{N!} \sum_{\wp}
\int d\rv \, \frac{\dens(\rv)}{N} \, \delta(\rv_1-\fv_{\wp(1)}(\rv)) \nonumber \\
&  \times\delta(\rv_2-\fv_{\wp(2)}(\rv)) \cdots \delta(\rv_N-\fv_{\wp(N)}(\rv)),
\label{eq_psi2}
\end{align}
%
and $\wp$ denote permutations of $\{1,\dots,N\}$.
The {\em co-motion functions} $\{\fv_i\}$ are highly non-local functionals of the 
density satisfying the equations \cite{SeiGorSav-PRA-07,GorSei-PCCP-10,ButDepGor-PRA-12} 
\be
\dens(\rv)d\rv = \dens(\fv_i(\rv))d\fv_i(\rv) \; ,
\label{eq_fi}
\ee
which are simply equivalent to state that the probability of finding particle ``1'' in the volume
element $d\rv$ around $\rv$ is the same as that of finding particle ``$i$'' in the volume 
element $d\fv_i(\rv)$ around $\fv_i(\rv)$. The $\{\fv_i\}$  also satisfy cyclic group properties that ensure the indistinguishability of the 
particles \cite{SeiGorSav-PRA-07,ButDepGor-PRA-12}. The SCP functional is then \cite{SeiGorSav-PRA-07,MirSeiGor-JCTC-12}
\be
V_{\rm int}^{\rm SCP}[\dens]= \frac{1}{2}\int d\rv\,\dens(\rv) \sum_{i=2}^N v_{\rm int}(\rv-\fv_i(\rv)).
\label{eq_VintSCP}
\ee
Notice that here we explicitly consider the possibility of anisotropic interactions depending on $\rv-\rv'$ (with $v_{\rm int}(\rv)=v_{\rm int}(-\rv)$) 
and not just on $|\rv-\rv'|$, such as the interaction between dipoles aligned by an external 
field \cite{Bar-RPP-08,LahMenSanLewPfa-RPP-09,BarDalPupZol-CR-12}. The same formal steps of Refs.~\cite{SeiGorSav-PRA-07,MalGor-PRL-12,MalMirCreReiGor-PRB-13} 
can be repeated for this more general kind of interaction, leading to the following exact equation for the functional 
derivative $v_{\rm SCP}([\dens];\rv)\equiv\frac{\delta V_{\rm int}^{\rm SCP}[\dens]}{\delta \dens(\rv)}$
\be
\nabla v_{\rm SCP}([\dens];\rv)=\sum_{i= 2}^N \nabla v_{\rm int}(\rv-\fv_i(\rv)),
\label{eq_vscp}
\ee
which has a clear physical meaning: the potential $v_{\rm SCP}(\rv)$ represents a force field equal to the net interaction felt by a particle at 
position $\rv$ due to the other $N-1$ particles. Our ``KS-SCP DFT'' approach consists in using $V_{\rm int}^{\rm SCP}[\dens]$ to approximate the 
mean-field plus exchange-correlation terms of the total energy functional or, equivalently, its functional derivative $v_{\rm SCP}([\dens];\rv)$  
of Eq.~\eqref{eq_vscp} to approximate the Hartree-exchange-correlation potential \cite{MalMirCreReiGor-PRB-13,MenMalGor-PRB-14},
\be
v_{\rm KS}([\dens];\rv)\simeq v_{\rm ext}(\rv) + v_{\rm SCP}([\dens];\rv) \; .
\label{eq_vscpks}
\ee
This way, both the kinetic energy and the many-body interactions are treated on the same footing in the self-consistent KS equations.

A few remarks are necessary on the kind of interactions $v_{\rm int}(\rv)$ for which the KS-SCP DFT can be applied. Several rigorous results 
are available for convex repulsive long-ranged interactions depending on $|\rv|$ only \cite{SeiGorSav-PRA-07,GorSei-PCCP-10,ColDepDiM-CJM-14,Pas-NL-13,Pas-JFA-13,GhoMoa-GFA-14,KimPas-SJMA-14,ColDiM-INC-13}. In general, 
for the SCP formalism to be physically useful, the interaction $v_{\rm int}(\rv)$ needs to be long-ranged, otherwise the SCP solution of 
Eq.~\eqref{eq_psi2} is just one of the many minimizers (and actually the one with maximum kinetic energy) for the interaction alone in a 
given density (see Ref.~\cite{RasSeiGor-PRB-11} for a discussion on contact interactions). The SCP functional is thus naturally very well 
suited for ionic gases in the strong-correlation regime, where it is expected to provide a large part of the total interaction 
energy \cite{MalGor-PRL-12,MalMirCreReiGor-PRB-13,MenMalGor-PRB-14}. Even more interesting is the case of general dipolar anisotropic 
interactions: the SCP functional combined with the KS kinetic energy (fermionic or bosonic) should be able to capture many of the 
interesting phenomena observed in the strong-correlation regime \cite{DeuCreRei-PRA-10,CreBruRei-PRL-10}. In this case, the SCP 
solution can be constructed from the dual Kantorovich formulation \cite{ButDepGor-PRA-12}, for which few results have started to 
appear recently \cite{MenLin-PRB-13,CheFriMen-JCTC-14}. 

\noindent{\it Applications to low-dimensional dipolar ultracold gases--} 
We consider $N$ ultracold bosonic or fermionic particles with dipole moment ${\bf d}$ in quasi-one- (Q1D) and 
quasi-two-dimensional (Q2D) geometries. 
We model these systems with the external harmonic potential 
$v_{\rm ext}(\rv)=\frac{1}{2}(\omega_x^2 x^2 + \omega_y^2 y^2 + \omega_z^2 z^2)$ in the cases where, respectively, 
$\omega_y$, $\omega_z \gg \omega_x$, and $\omega_z\gg\omega_x=\omega_y\equiv \omega_\bot$ (effective Hartree 
units are used throughout the paper). For these geometries, assuming that all the dipoles are oriented in the same direction
due to the action of some external field, one can derive effective Q1D and Q2D dipole-dipole interaction potentials 
(see e.g. Refs. \cite{DeuCreRei-PRA-10} and \cite{CreBruRei-PRL-10} for the explicit expressions) by integrating out the harmonic 
motion along the very strongly confined directions from the three-dimensional potential $v_{dd}(\rv)=d^2(1-3\cos^2 \theta_{\rm rd})/r^3$, 
where $\theta_{\rm rd}$ is the angle between the dipole moment and the relative position between two particles. Since the anisotropic 
interaction requires the development of a dedicated dual Kantorovich algorithm that will be the object of future work, here for the 
Q2D dipolar systems we restrict ourselves to the case in which the dipoles are perpendicular to their plane of motion and their 
interaction is purely repulsive and isotropic \cite{CreBruRei-PRL-10}.

In order to perform practical calculations with the KS-SCP approach, one proceeds as follows. First, for a given density distribution, 
the co-motion functions $\{\fv_i\}$ are obtained by solving Eqs.~(\ref{eq_fi}) and performing an angular 
minimization \cite{MalMirCreReiGor-PRB-13,MenMalGor-PRB-14}. The one-body potential $v_{\rm SCP}(\rv)$ can then be obtained by integrating 
Eq.~(\ref{eq_vscp}). Finally, one solves the Kohn-Sham equations using the approximation of Eq.~(\ref{eq_vscpks}) in order to obtain 
a new density. The process is then repeated self-consistently until convergence is achieved.

\begin{figure}[t]
\includegraphics[width=7.0cm]{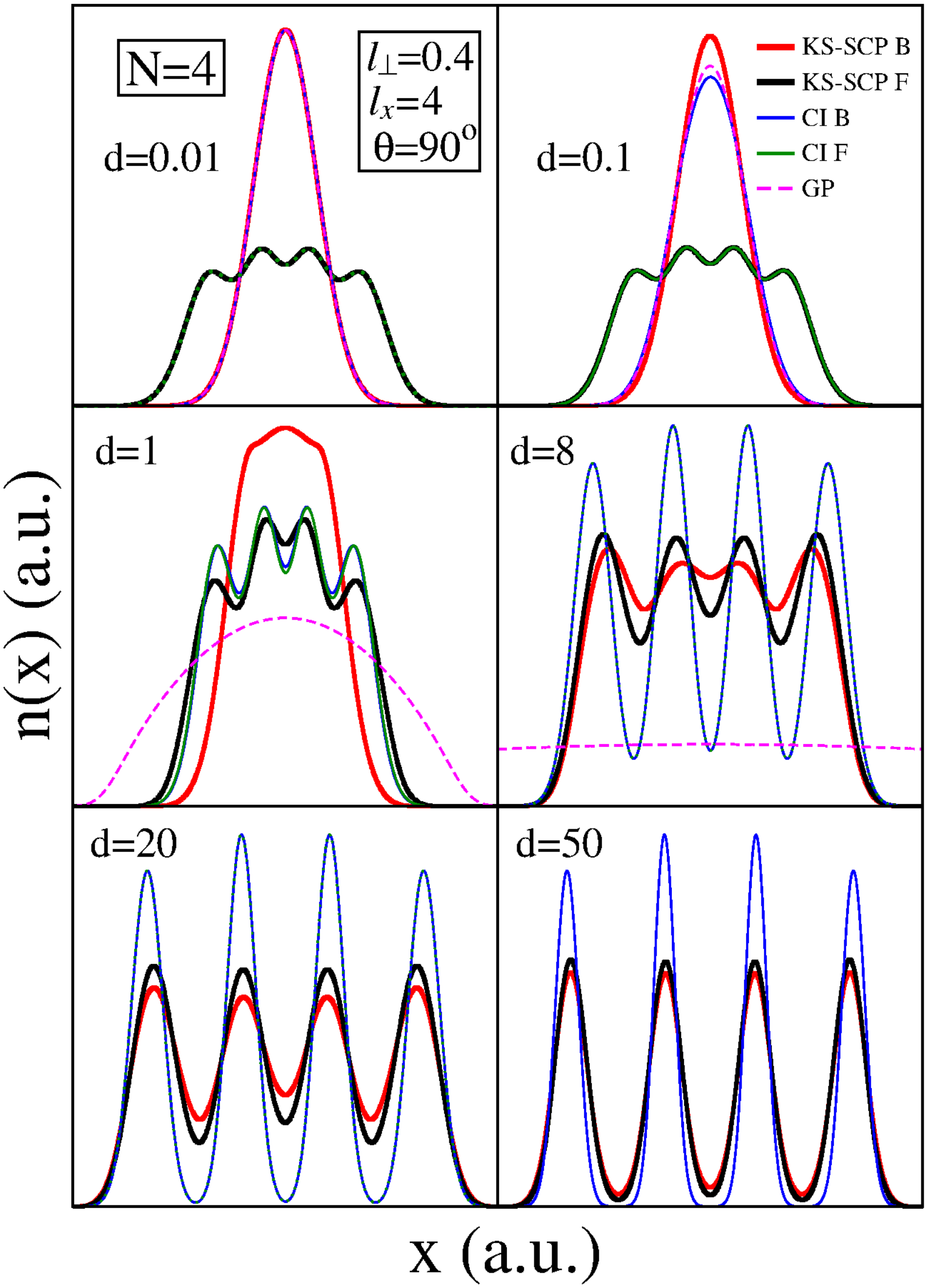}
\caption{(color online) Densities for $N=4$ dipoles in a Q1D geometry ($l_x=10l_\bot$) as a 
function of the dipole moment $d$, which is assumed to be oriented perpendicularly to the direction of motion. 
The different results correspond to the KS-SCP DFT (red for bosons (B), black for fermions (F)), CI (blue for 
bosons, green for fermions), and GP (magenta) approaches.}
\label{fig_dens_N4d}
\end{figure}

To illustrate the kind of accuracy that can be obtained with the KS-SCP formalism, we chose a case for 
which full CI calculations can be also performed:  in Fig.~\ref{fig_dens_N4d} 
we report the KS-SCP densities for a system with four bosonic and fermionic dipoles in a 
Q1D trap. To model this system we have chosen the oscillator lengths in the longitudinal
and perpendicular directions $l_x=\sqrt{\hbar/(m\omega_x)}$ and $l_{\bot}=\sqrt{\hbar/(m\omega_\bot)}$
such that $l_x=10 l_\bot$. Also, we have assumed the dipoles to be oriented perpendicularly to their 
direction of motion, that is $\theta_{\rm rd}\equiv \Theta=90^{\circ}$. For comparison, we also show the 
results obtained from the mean-field GP approach for the bosonic case. When $d$ is very small 
(0.01 and 0.1) the system is in the weakly-interacting regime: we see good qualitative agreement of the 
KS-SCP results with the CI for fermions, and also with the GP calculations for bosons, reflecting the fact that the correlation effects are 
negligible. As $d$ is further increased, however, the picture changes qualitatively. For $d=1$ the CI bosonic
results show a density structure characteristic of the so-called fermionized regime \cite{DeuCreRei-PRA-10}, 
with two tall central peaks and two shorter lateral ones, indicating that the bosons feel the
infinitely strong short-ranged part of the interaction and behave as non-interacting spinless fermions. 
The zeroth-order KS-SCP approach can only describe this phenomenon in a qualitative way, barely displaying 
two incipient lateral peaks but being not able to reproduce the central structure for the chosen parameters. 
This is due to the large underestimation of the kinetic correlation energy: in the SCP state of Eq.~\eqref{eq_psi2}, the particles are located in 
their strictly-correlated positions, minimizing the interaction energy for the given 
density, but without increasing substantially the kinetic energy, which is obtained from the KS construction. The resulting SCP potential of Eq.~\eqref{eq_vscp} is then too small, since the particle are too far from each other. Still, one can 
see that the KS-SCP results are much better than those obtained from the GP approach: indeed, 
the latter yields a Thomas-Fermi-like density profile lacking of any structure. Finally, for strong enough 
values of the dipole moment ($d=8$, 20 and 50), the system enters the localized regime, where the CI densities 
show a characteristic profile with four clearly marked peaks corresponding to the localization of the 
density \cite{DeuCreRei-PRA-10} due to the strong long-range repulsion between the particles. One can see how 
the KS-SCP densities show this structure as well, and that they become closer to the exact ones as the strength 
of the interaction increases. The capacity of the KS SCP approach for going beyond the mean-field description is 
apparent from the case $d=8$, where the GP density illustrates the total neglect of correlation effects (for larger $d$ we could not even get converged GP results within our grid). It 
is also remarkable that the same approximate method is able to span (even if only qualitatively) a wide range 
of different correlation regimes. Moreover, as any DFT approach, the formalism is amenable to corrections: 
exchange effects, for example, could be included in the SCP functional in order to capture the fermionized regime. 
Notice that while full CI calculations almost reached the maximum number of particles that can be treated, the 
computational cost of the KS-SCP method in one dimension is similar to that of the usual LDA.

\begin{figure}[t]
\includegraphics[width=8.5cm]{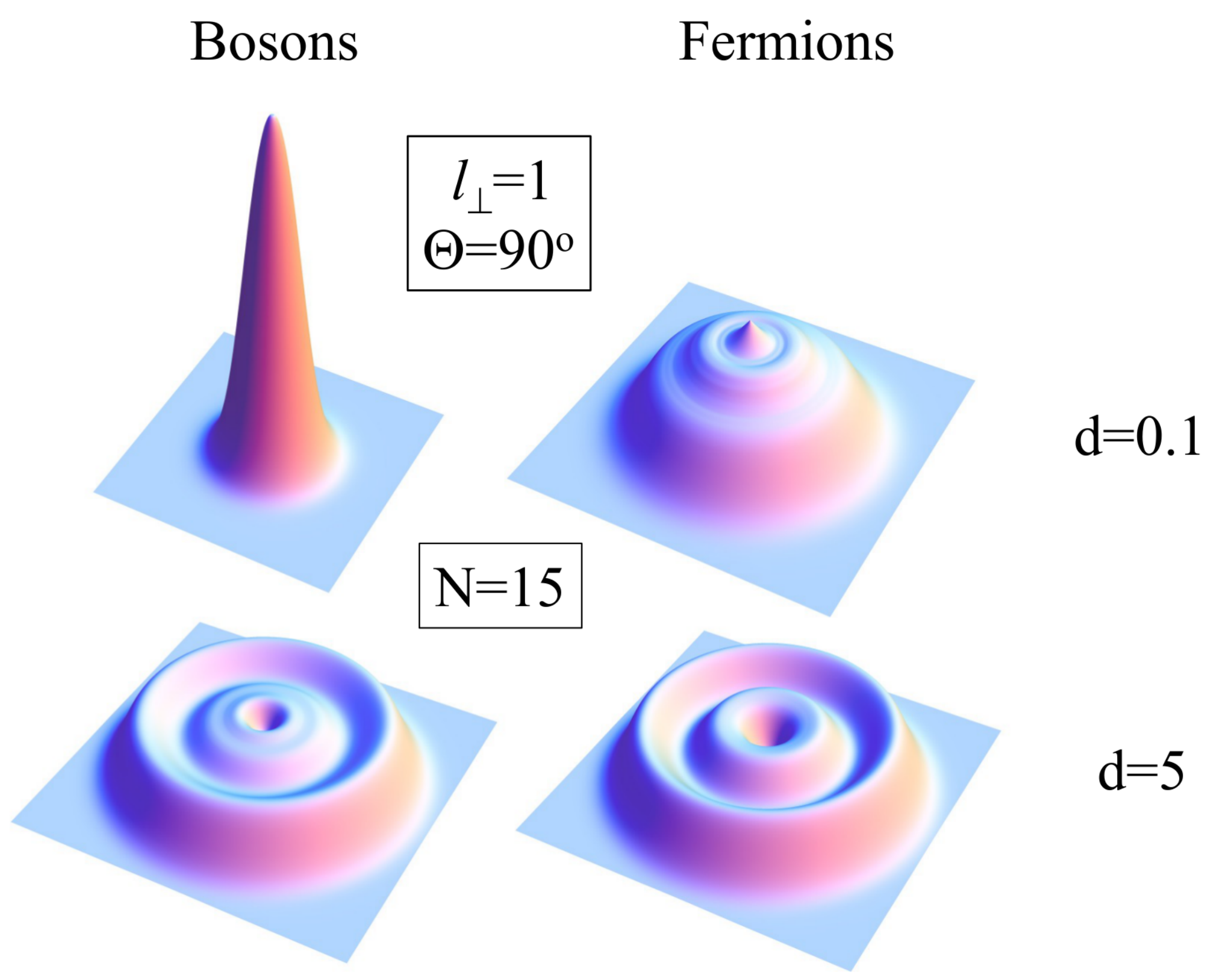}
\caption{(color online) KS-SCP densities for bosonic and fermionic systems in
  a Q2D geometry for $N=15$ in the weakly- and strongly-correlated regimes.}
\label{fig2DN15}
\end{figure}

In Fig.~\ref{fig2DN15} we show the KS-SCP densities for Q2D systems with $N=15$ particles, considering 
the dipole moments to be perpendicularly aligned with the plane of motion, and setting $10 l_z = l_{\bot} = l_x = l_y = 1$. Similarly 
as in the Q1D case of Fig.~\ref{fig_dens_N4d}, one can see how the KS SCP method is able to span the range from 
the weakly-interacting regime (corresponding to Bose-Einstein condensation for bosons and to the shell structure for fermions), 
to the strongly-correlated one. In this latter case, one can observe the formation of two characteristic concentric rings, 
which by integration of the density reveal the presence of 6 and 9 particles on average. Again, these calculations are out 
of reach of full CI, and also difficult with QMC.


\noindent{\it Conclusions -- }In this Letter we have suggested an exchange-correlation functional for the application
of Kohn-Sham DFT to ultracold bosonic and fermionic dipolar and ionic gases. This functional is based on the 
exact strong-coupling limit for a given density $\dens(\rv)$, and can be used to treat long-ranged interactions
in the strongly-correlated regime. The functional can be improved with rigorous 
corrections \cite{GorVigSei-JCTC-09,GorSeiVig-PRL-09} and one should expect that inclusion of higher-order 
terms would improve the performance for intermediate regimes. 

The results show the potential of strictly-correlated-based density functionals for the study of ultracold 
dipolar and ionic gases, where only little progress has been done in the application of density-functional 
methods \cite{FanEng-PRA-11} compared to ultracold atomic systems with short-ranged interactions. The unique 
properties and high tunability of the dipole-dipole interaction have been shown to be potentially useful for the 
study of many interesting phenomena, as well as for several practical applications. Their investigation with 
density-functional methods might open new possibilities, especially when generalized to the time domain 
(which is the object of ongoing work). The presented method can be applied to systems with 
other external potentials, such as the interesting case of 1D optical lattices, as well as generalized 
to different interactions and higher dimensionalities. In this sense, one interesting possibility could
be the recently discussed quadrupolar interactions, which may lead to intriguing new quantum phases 
(see \cite{LahMikMat-arx-14,BhoMatZhaYelLem-PRL-13,HuaLahMat-arx-13}). These topics will be the subject 
of future works.

\noindent{\it Acknowledgments --}
We thank D.~Karlsson, G.~Kavoulakis, I.~Tokatly, C.~Verdozzi and G.~Vignale for fruitful discussions. 
We acknowledge support from the Netherlands Organization for Scientific Research (NWO) through a 
Vidi grant, from a Marie Curie Intra European Fellowship within the 7th European Community 
Framework Programme, from the Swedish Research Council, and from the Nanometer Structure
Consortium at Lund University.

\bibliography{biblio}

\begin{thebibliography}{60}
\expandafter\ifx\csname natexlab\endcsname\relax\def\natexlab#1{#1}\fi
\expandafter\ifx\csname bibnamefont\endcsname\relax
  \def\bibnamefont#1{#1}\fi
\expandafter\ifx\csname bibfnamefont\endcsname\relax
  \def\bibfnamefont#1{#1}\fi
\expandafter\ifx\csname citenamefont\endcsname\relax
  \def\citenamefont#1{#1}\fi
\expandafter\ifx\csname url\endcsname\relax
  \def\url#1{\texttt{#1}}\fi
\expandafter\ifx\csname urlprefix\endcsname\relax\def\urlprefix{URL }\fi
\providecommand{\bibinfo}[2]{#2}
\providecommand{\eprint}[2][]{\url{#2}}

\bibitem[{\citenamefont{Cohen et~al.}(2012)\citenamefont{Cohen, Mori-S\'anchez,
  and Yang}}]{CohMorYan-CR-12}
\bibinfo{author}{\bibfnamefont{A.~J.} \bibnamefont{Cohen}},
  \bibinfo{author}{\bibfnamefont{P.}~\bibnamefont{Mori-S\'anchez}},
  \bibnamefont{and} \bibinfo{author}{\bibfnamefont{W.}~\bibnamefont{Yang}},
  \bibinfo{journal}{Chem. Rev.} \textbf{\bibinfo{volume}{112}},
  \bibinfo{pages}{289} (\bibinfo{year}{2012}).

\bibitem[{\citenamefont{Neugebauer and Hickel}(2013)}]{NeuTil-13}
\bibinfo{author}{\bibfnamefont{J.}~\bibnamefont{Neugebauer}} \bibnamefont{and}
  \bibinfo{author}{\bibfnamefont{T.}~\bibnamefont{Hickel}},
  \bibinfo{journal}{Wiley Interdisciplinary Reviews: Computational Molecular
  Science} \textbf{\bibinfo{volume}{{3}}}, \bibinfo{pages}{438}
  (\bibinfo{year}{2013}).

\bibitem[{\citenamefont{Reimann and Manninen}(2002)}]{ReiMan-RMP-02}
\bibinfo{author}{\bibfnamefont{S.~M.} \bibnamefont{Reimann}} \bibnamefont{and}
  \bibinfo{author}{\bibfnamefont{M.}~\bibnamefont{Manninen}},
  \bibinfo{journal}{Rev. Mod. Phys.} \textbf{\bibinfo{volume}{{74}}},
  \bibinfo{pages}{1283} (\bibinfo{year}{2002}).

\bibitem[{\citenamefont{Hohenberg and Kohn}(1964)}]{HohKoh-PR-64}
\bibinfo{author}{\bibfnamefont{P.}~\bibnamefont{Hohenberg}} \bibnamefont{and}
  \bibinfo{author}{\bibfnamefont{W.}~\bibnamefont{Kohn}},
  \bibinfo{journal}{Phys. Rev.} \textbf{\bibinfo{volume}{{136}}},
  \bibinfo{pages}{B 864} (\bibinfo{year}{1964}).

\bibitem[{\citenamefont{Kohn and Sham}(1965)}]{KohSha-PR-65}
\bibinfo{author}{\bibfnamefont{W.}~\bibnamefont{Kohn}} \bibnamefont{and}
  \bibinfo{author}{\bibfnamefont{L.~J.} \bibnamefont{Sham}},
  \bibinfo{journal}{Phys. Rev. A} \textbf{\bibinfo{volume}{140}},
  \bibinfo{pages}{1133} (\bibinfo{year}{1965}).

\bibitem[{\citenamefont{Bloch et~al.}(2008)\citenamefont{Bloch, Dalibard, and
  Zwerger}}]{BloDalZwe-RMP-08}
\bibinfo{author}{\bibfnamefont{I.}~\bibnamefont{Bloch}},
  \bibinfo{author}{\bibfnamefont{J.}~\bibnamefont{Dalibard}}, \bibnamefont{and}
  \bibinfo{author}{\bibfnamefont{W.}~\bibnamefont{Zwerger}},
  \bibinfo{journal}{Rev. Mod. Phys.} \textbf{\bibinfo{volume}{{80}}},
  \bibinfo{pages}{885} (\bibinfo{year}{2008}).

\bibitem[{\citenamefont{Minguzzi et~al.}(2004)\citenamefont{Minguzzi, Succi,
  Toschi, Tosi, and Vignolo}}]{MinSucTosTosVig-PR-04}
\bibinfo{author}{\bibfnamefont{A.}~\bibnamefont{Minguzzi}},
  \bibinfo{author}{\bibfnamefont{S.}~\bibnamefont{Succi}},
  \bibinfo{author}{\bibfnamefont{F.}~\bibnamefont{Toschi}},
  \bibinfo{author}{\bibfnamefont{M.}~\bibnamefont{Tosi}}, \bibnamefont{and}
  \bibinfo{author}{\bibfnamefont{P.}~\bibnamefont{Vignolo}},
  \bibinfo{journal}{Phys. Rep.} \textbf{\bibinfo{volume}{{395}}},
  \bibinfo{pages}{223} (\bibinfo{year}{2004}).

\bibitem[{\citenamefont{S.~Giorgini and Stringari}(2008)}]{GioPitStri-RMP-08}
\bibinfo{author}{\bibfnamefont{L.~P.~P.} \bibnamefont{S.~Giorgini}}
  \bibnamefont{and}
  \bibinfo{author}{\bibfnamefont{S.}~\bibnamefont{Stringari}},
  \bibinfo{journal}{Rev. Mod. Phys.} \textbf{\bibinfo{volume}{{80}}},
  \bibinfo{pages}{1215} (\bibinfo{year}{2008}).

\bibitem[{\citenamefont{Rontani et~al.}(2006)\citenamefont{Rontani, Cavazzoni,
  Bellucci, and Goldoni}}]{RonCavBelGol-JCP-06}
\bibinfo{author}{\bibfnamefont{M.}~\bibnamefont{Rontani}},
  \bibinfo{author}{\bibfnamefont{C.}~\bibnamefont{Cavazzoni}},
  \bibinfo{author}{\bibfnamefont{D.}~\bibnamefont{Bellucci}}, \bibnamefont{and}
  \bibinfo{author}{\bibfnamefont{G.}~\bibnamefont{Goldoni}},
  \bibinfo{journal}{J. Chem. Phys.} \textbf{\bibinfo{volume}{{124}}},
  \bibinfo{pages}{124102} (\bibinfo{year}{2006}).

\bibitem[{\citenamefont{Waltersson et~al.}(2013)\citenamefont{Waltersson,
  Wessl\'en, and Lindroth}}]{WalWesLin-PRB-13}
\bibinfo{author}{\bibfnamefont{E.}~\bibnamefont{Waltersson}},
  \bibinfo{author}{\bibfnamefont{C.~J.} \bibnamefont{Wessl\'en}},
  \bibnamefont{and} \bibinfo{author}{\bibfnamefont{E.}~\bibnamefont{Lindroth}},
  \bibinfo{journal}{Phys. Rev. B} \textbf{\bibinfo{volume}{87}},
  \bibinfo{pages}{035112} (\bibinfo{year}{2013}).

\bibitem[{\citenamefont{Ghosal et~al.}(2006)\citenamefont{Ghosal, Guclu,
  Umrigar, Ullmo, and Baranger}}]{GhoGucUmrUllBar-NP-06}
\bibinfo{author}{\bibfnamefont{A.}~\bibnamefont{Ghosal}},
  \bibinfo{author}{\bibfnamefont{A.~D.} \bibnamefont{Guclu}},
  \bibinfo{author}{\bibfnamefont{C.~J.} \bibnamefont{Umrigar}},
  \bibinfo{author}{\bibfnamefont{D.}~\bibnamefont{Ullmo}}, \bibnamefont{and}
  \bibinfo{author}{\bibfnamefont{H.~U.} \bibnamefont{Baranger}},
  \bibinfo{journal}{Nature Phys.} \textbf{\bibinfo{volume}{2}},
  \bibinfo{pages}{336} (\bibinfo{year}{2006}).

\bibitem[{\citenamefont{Ghosal et~al.}(2007)\citenamefont{Ghosal, Guclu,
  Umrigar, Ullmo, and Baranger}}]{GhoGucUmrUllBar-PRB-07}
\bibinfo{author}{\bibfnamefont{A.}~\bibnamefont{Ghosal}},
  \bibinfo{author}{\bibfnamefont{A.~D.} \bibnamefont{Guclu}},
  \bibinfo{author}{\bibfnamefont{C.~J.} \bibnamefont{Umrigar}},
  \bibinfo{author}{\bibfnamefont{D.}~\bibnamefont{Ullmo}}, \bibnamefont{and}
  \bibinfo{author}{\bibfnamefont{H.~U.} \bibnamefont{Baranger}},
  \bibinfo{journal}{Phys. Rev. B} \textbf{\bibinfo{volume}{76}},
  \bibinfo{pages}{085341} (\bibinfo{year}{2007}).

\bibitem[{\citenamefont{Nunes}(1999)}]{Nun-JPB-99}
\bibinfo{author}{\bibfnamefont{G.~S.} \bibnamefont{Nunes}},
  \bibinfo{journal}{J. Phys. B: At. Mol. Opt. Phys.}
  \textbf{\bibinfo{volume}{{32}}}, \bibinfo{pages}{4293}
  (\bibinfo{year}{1999}).

\bibitem[{\citenamefont{Kim and Zubarev}(2003)}]{KimZub-PRA-03}
\bibinfo{author}{\bibfnamefont{Y.~E.} \bibnamefont{Kim}} \bibnamefont{and}
  \bibinfo{author}{\bibfnamefont{A.~L.} \bibnamefont{Zubarev}},
  \bibinfo{journal}{Phys. Rev. A} \textbf{\bibinfo{volume}{{67}}},
  \bibinfo{pages}{015602} (\bibinfo{year}{2003}).

\bibitem[{\citenamefont{Brand}(2004)}]{Bra-JPB-04}
\bibinfo{author}{\bibfnamefont{J.}~\bibnamefont{Brand}}, \bibinfo{journal}{J.
  Phys. B: At. Mol. Opt. Phys.} \textbf{\bibinfo{volume}{{37}}},
  \bibinfo{pages}{287} (\bibinfo{year}{2004}).

\bibitem[{\citenamefont{Hao and Chen}(2009)}]{HaoChe-PRA-09}
\bibinfo{author}{\bibfnamefont{Y.}~\bibnamefont{Hao}} \bibnamefont{and}
  \bibinfo{author}{\bibfnamefont{S.}~\bibnamefont{Chen}},
  \bibinfo{journal}{Phys. Rev. A} \textbf{\bibinfo{volume}{{80}}},
  \bibinfo{pages}{043608} (\bibinfo{year}{2009}).

\bibitem[{\citenamefont{Ma et~al.}(2012)\citenamefont{Ma, Pilati, Troyer, and
  Dai}}]{PingPilTroDai-Nat-12}
\bibinfo{author}{\bibfnamefont{P.~N.} \bibnamefont{Ma}},
  \bibinfo{author}{\bibfnamefont{S.}~\bibnamefont{Pilati}},
  \bibinfo{author}{\bibfnamefont{M.}~\bibnamefont{Troyer}}, \bibnamefont{and}
  \bibinfo{author}{\bibfnamefont{X.}~\bibnamefont{Dai}},
  \bibinfo{journal}{Nature Phys.} \textbf{\bibinfo{volume}{{8}}},
  \bibinfo{pages}{601} (\bibinfo{year}{2012}).

\bibitem[{\citenamefont{Casula et~al.}(2006)\citenamefont{Casula, Sorella, and
  Senatore}}]{CasSorSen-PRB-06}
\bibinfo{author}{\bibfnamefont{M.}~\bibnamefont{Casula}},
  \bibinfo{author}{\bibfnamefont{S.}~\bibnamefont{Sorella}}, \bibnamefont{and}
  \bibinfo{author}{\bibfnamefont{G.}~\bibnamefont{Senatore}},
  \bibinfo{journal}{Phys. Rev. B} \textbf{\bibinfo{volume}{{74}}},
  \bibinfo{pages}{245427} (\bibinfo{year}{2006}).

\bibitem[{\citenamefont{Lima et~al.}(2003)\citenamefont{Lima, Silva, Oliveira,
  and Capelle}}]{LimSilOliCap-PRL-03}
\bibinfo{author}{\bibfnamefont{N.~A.} \bibnamefont{Lima}},
  \bibinfo{author}{\bibfnamefont{M.~F.} \bibnamefont{Silva}},
  \bibinfo{author}{\bibfnamefont{L.~N.} \bibnamefont{Oliveira}},
  \bibnamefont{and} \bibinfo{author}{\bibfnamefont{K.}~\bibnamefont{Capelle}},
  \bibinfo{journal}{Phys. Rev. Lett.} \textbf{\bibinfo{volume}{{90}}},
  \bibinfo{pages}{146402} (\bibinfo{year}{2003}).

\bibitem[{\citenamefont{Wang et~al.}(2012)\citenamefont{Wang, Hao, and
  Zhang}}]{WanHaoZha-PRA-12}
\bibinfo{author}{\bibfnamefont{H.}~\bibnamefont{Wang}},
  \bibinfo{author}{\bibfnamefont{Y.}~\bibnamefont{Hao}}, \bibnamefont{and}
  \bibinfo{author}{\bibfnamefont{Y.}~\bibnamefont{Zhang}},
  \bibinfo{journal}{Phys. Rev. A} \textbf{\bibinfo{volume}{{85}}},
  \bibinfo{pages}{053630} (\bibinfo{year}{2012}).

\bibitem[{\citenamefont{Wang and Zhang}(2013)}]{WanZha-PRA-13}
\bibinfo{author}{\bibfnamefont{H.}~\bibnamefont{Wang}} \bibnamefont{and}
  \bibinfo{author}{\bibfnamefont{Y.}~\bibnamefont{Zhang}},
  \bibinfo{journal}{Phys. Rev. A} \textbf{\bibinfo{volume}{{88}}},
  \bibinfo{pages}{023626} (\bibinfo{year}{2013}).

\bibitem[{\citenamefont{Baranov}(2008)}]{Bar-RPP-08}
\bibinfo{author}{\bibfnamefont{M.~A.} \bibnamefont{Baranov}},
  \bibinfo{journal}{Physics Reports} \textbf{\bibinfo{volume}{{464}}},
  \bibinfo{pages}{71} (\bibinfo{year}{2008}).

\bibitem[{\citenamefont{Lahaye et~al.}(2009)\citenamefont{Lahaye, Menotti,
  Santos, Lewenstein, and Pfau}}]{LahMenSanLewPfa-RPP-09}
\bibinfo{author}{\bibfnamefont{T.}~\bibnamefont{Lahaye}},
  \bibinfo{author}{\bibfnamefont{C.}~\bibnamefont{Menotti}},
  \bibinfo{author}{\bibfnamefont{L.}~\bibnamefont{Santos}},
  \bibinfo{author}{\bibfnamefont{M.}~\bibnamefont{Lewenstein}},
  \bibnamefont{and} \bibinfo{author}{\bibfnamefont{T.}~\bibnamefont{Pfau}},
  \bibinfo{journal}{Rep. Prog. Phys.} \textbf{\bibinfo{volume}{{72}}},
  \bibinfo{pages}{126401} (\bibinfo{year}{2009}).

\bibitem[{\citenamefont{Baranov et~al.}(2012)\citenamefont{Baranov, Dalmonte,
  Pupillo, and Zoller}}]{BarDalPupZol-CR-12}
\bibinfo{author}{\bibfnamefont{M.~A.} \bibnamefont{Baranov}},
  \bibinfo{author}{\bibfnamefont{M.}~\bibnamefont{Dalmonte}},
  \bibinfo{author}{\bibfnamefont{G.}~\bibnamefont{Pupillo}}, \bibnamefont{and}
  \bibinfo{author}{\bibfnamefont{P.}~\bibnamefont{Zoller}},
  \bibinfo{journal}{Chem. Rev.} \textbf{\bibinfo{volume}{{112}}},
  \bibinfo{pages}{5012} (\bibinfo{year}{2012}).

\bibitem[{\citenamefont{Ni et~al.}(2008)\citenamefont{Ni, \"Ospelkaus,
  de~Miranda, Pe'er, Neyenhuis, , Zirbel, Kotochigova, Julienne, Jin1
  et~al.}}]{Ni-Sci-08}
\bibinfo{author}{\bibfnamefont{K.-K.} \bibnamefont{Ni}},
  \bibinfo{author}{\bibfnamefont{S.}~\bibnamefont{\"Ospelkaus}},
  \bibinfo{author}{\bibfnamefont{M.~H.~G.} \bibnamefont{de~Miranda}},
  \bibinfo{author}{\bibfnamefont{A.}~\bibnamefont{Pe'er}},
  \bibinfo{author}{\bibfnamefont{B.}~\bibnamefont{Neyenhuis}}, ,
  \bibinfo{author}{\bibfnamefont{J.~J.} \bibnamefont{Zirbel}},
  \bibinfo{author}{\bibfnamefont{S.}~\bibnamefont{Kotochigova}},
  \bibinfo{author}{\bibfnamefont{P.~S.} \bibnamefont{Julienne}},
  \bibinfo{author}{\bibfnamefont{D.~S.} \bibnamefont{Jin1}},
  \bibnamefont{et~al.}, \bibinfo{journal}{Science}
  \textbf{\bibinfo{volume}{{322}}}, \bibinfo{pages}{231}
  (\bibinfo{year}{2008}).

\bibitem[{\citenamefont{Ospelkaus et~al.}(2010)\citenamefont{Ospelkaus, Ni,
  Wang, de~Miranda, Neyenhuis, Qu\'em\'ener, Julienne, Bohn, Jin, and
  Ye}}]{Osp-Sci-10}
\bibinfo{author}{\bibfnamefont{S.}~\bibnamefont{Ospelkaus}},
  \bibinfo{author}{\bibfnamefont{K.-K.} \bibnamefont{Ni}},
  \bibinfo{author}{\bibfnamefont{D.}~\bibnamefont{Wang}},
  \bibinfo{author}{\bibfnamefont{M.~H.~G.} \bibnamefont{de~Miranda}},
  \bibinfo{author}{\bibfnamefont{B.}~\bibnamefont{Neyenhuis}},
  \bibinfo{author}{\bibfnamefont{G.}~\bibnamefont{Qu\'em\'ener}},
  \bibinfo{author}{\bibfnamefont{P.~S.} \bibnamefont{Julienne}},
  \bibinfo{author}{\bibfnamefont{J.~L.} \bibnamefont{Bohn}},
  \bibinfo{author}{\bibfnamefont{D.~S.} \bibnamefont{Jin}}, \bibnamefont{and}
  \bibinfo{author}{\bibfnamefont{J.}~\bibnamefont{Ye}},
  \bibinfo{journal}{Science} \textbf{\bibinfo{volume}{{327}}},
  \bibinfo{pages}{853} (\bibinfo{year}{2010}).

\bibitem[{\citenamefont{Astrakharchik et~al.}(2008)\citenamefont{Astrakharchik,
  Morigi, DeChiara, and Boronat}}]{AstMorChiBor-PRA-08}
\bibinfo{author}{\bibfnamefont{G.~E.} \bibnamefont{Astrakharchik}},
  \bibinfo{author}{\bibfnamefont{G.}~\bibnamefont{Morigi}},
  \bibinfo{author}{\bibfnamefont{G.}~\bibnamefont{DeChiara}}, \bibnamefont{and}
  \bibinfo{author}{\bibfnamefont{J.}~\bibnamefont{Boronat}},
  \bibinfo{journal}{Phys. Rev. A} \textbf{\bibinfo{volume}{{78}}},
  \bibinfo{pages}{063622} (\bibinfo{year}{2008}).

\bibitem[{\citenamefont{Zinner et~al.}(2011)\citenamefont{Zinner, Wunsch,
  Mekhov, Huang, Wang, and Demler}}]{ZinWunMekHuaWanDem-PRA-11}
\bibinfo{author}{\bibfnamefont{N.~T.} \bibnamefont{Zinner}},
  \bibinfo{author}{\bibfnamefont{B.}~\bibnamefont{Wunsch}},
  \bibinfo{author}{\bibfnamefont{I.}~\bibnamefont{Mekhov}},
  \bibinfo{author}{\bibfnamefont{S.-J.} \bibnamefont{Huang}},
  \bibinfo{author}{\bibfnamefont{D.}~\bibnamefont{Wang}}, \bibnamefont{and}
  \bibinfo{author}{\bibfnamefont{E.}~\bibnamefont{Demler}},
  \bibinfo{journal}{Phys. Rev. A} \textbf{\bibinfo{volume}{{84}}},
  \bibinfo{pages}{063606} (\bibinfo{year}{2011}).

\bibitem[{\citenamefont{Deuretzbacher et~al.}(2010)\citenamefont{Deuretzbacher,
  Cremon, and Reimann}}]{DeuCreRei-PRA-10}
\bibinfo{author}{\bibfnamefont{F.}~\bibnamefont{Deuretzbacher}},
  \bibinfo{author}{\bibfnamefont{J.~C.} \bibnamefont{Cremon}},
  \bibnamefont{and} \bibinfo{author}{\bibfnamefont{S.~M.}
  \bibnamefont{Reimann}}, \bibinfo{journal}{Phys. Rev. A.}
  \textbf{\bibinfo{volume}{{81}}}, \bibinfo{pages}{063616}
  (\bibinfo{year}{2010}).

\bibitem[{\citenamefont{Dalmonte et~al.}(2010)\citenamefont{Dalmonte, Pupillo,
  and Zoller}}]{DalPupZol-PRL-10}
\bibinfo{author}{\bibfnamefont{M.}~\bibnamefont{Dalmonte}},
  \bibinfo{author}{\bibfnamefont{G.}~\bibnamefont{Pupillo}}, \bibnamefont{and}
  \bibinfo{author}{\bibfnamefont{P.}~\bibnamefont{Zoller}},
  \bibinfo{journal}{Phys. Rev. Lett.} \textbf{\bibinfo{volume}{{105}}},
  \bibinfo{pages}{140401} (\bibinfo{year}{2010}).

\bibitem[{\citenamefont{Citro et~al.}(2007)\citenamefont{Citro, Orignac, Palo,
  and Chiofalo}}]{CitOriPalChi-PRA-07}
\bibinfo{author}{\bibfnamefont{R.}~\bibnamefont{Citro}},
  \bibinfo{author}{\bibfnamefont{E.}~\bibnamefont{Orignac}},
  \bibinfo{author}{\bibfnamefont{S.~D.} \bibnamefont{Palo}}, \bibnamefont{and}
  \bibinfo{author}{\bibfnamefont{M.}~\bibnamefont{Chiofalo}},
  \bibinfo{journal}{Phys. Rev. A} \textbf{\bibinfo{volume}{{75}}},
  \bibinfo{pages}{051602(R)} (\bibinfo{year}{2007}).

\bibitem[{\citenamefont{Wunsch et~al.}(2011)\citenamefont{Wunsch, Zinner,
  Mekov, Huang, Wang, and Demler}}]{WunZinMekHuaWanDem-PRL-11}
\bibinfo{author}{\bibfnamefont{B.}~\bibnamefont{Wunsch}},
  \bibinfo{author}{\bibfnamefont{N.}~\bibnamefont{Zinner}},
  \bibinfo{author}{\bibfnamefont{I.}~\bibnamefont{Mekov}},
  \bibinfo{author}{\bibfnamefont{S.-J.} \bibnamefont{Huang}},
  \bibinfo{author}{\bibfnamefont{D.-W.} \bibnamefont{Wang}}, \bibnamefont{and}
  \bibinfo{author}{\bibfnamefont{E.}~\bibnamefont{Demler}},
  \bibinfo{journal}{Phys. Rev. Lett.} \textbf{\bibinfo{volume}{{107}}},
  \bibinfo{pages}{073201} (\bibinfo{year}{2011}).

\bibitem[{\citenamefont{Gori-Giorgi
  et~al.}(2009{\natexlab{a}})\citenamefont{Gori-Giorgi, Seidl, and
  Vignale}}]{GorSeiVig-PRL-09}
\bibinfo{author}{\bibfnamefont{P.}~\bibnamefont{Gori-Giorgi}},
  \bibinfo{author}{\bibfnamefont{M.}~\bibnamefont{Seidl}}, \bibnamefont{and}
  \bibinfo{author}{\bibfnamefont{G.}~\bibnamefont{Vignale}},
  \bibinfo{journal}{Phys. Rev. Lett.} \textbf{\bibinfo{volume}{{103}}},
  \bibinfo{pages}{166402} (\bibinfo{year}{2009}{\natexlab{a}}).

\bibitem[{\citenamefont{Gori-Giorgi and Seidl}(2010)}]{GorSei-PCCP-10}
\bibinfo{author}{\bibfnamefont{P.}~\bibnamefont{Gori-Giorgi}} \bibnamefont{and}
  \bibinfo{author}{\bibfnamefont{M.}~\bibnamefont{Seidl}},
  \bibinfo{journal}{Phys. Chem. Chem. Phys.} \textbf{\bibinfo{volume}{{12}}},
  \bibinfo{pages}{14405} (\bibinfo{year}{2010}).

\bibitem[{\citenamefont{Malet and Gori-Giorgi}(2012)}]{MalGor-PRL-12}
\bibinfo{author}{\bibfnamefont{F.}~\bibnamefont{Malet}} \bibnamefont{and}
  \bibinfo{author}{\bibfnamefont{P.}~\bibnamefont{Gori-Giorgi}},
  \bibinfo{journal}{Phys. Rev. Lett.} \textbf{\bibinfo{volume}{{109}}},
  \bibinfo{pages}{246402} (\bibinfo{year}{2012}).

\bibitem[{\citenamefont{Malet et~al.}(2013)\citenamefont{Malet, Mirtschink,
  Cremon, Reimann, and Gori-Giorgi}}]{MalMirCreReiGor-PRB-13}
\bibinfo{author}{\bibfnamefont{F.}~\bibnamefont{Malet}},
  \bibinfo{author}{\bibfnamefont{A.}~\bibnamefont{Mirtschink}},
  \bibinfo{author}{\bibfnamefont{J.~C.} \bibnamefont{Cremon}},
  \bibinfo{author}{\bibfnamefont{S.~M.} \bibnamefont{Reimann}},
  \bibnamefont{and}
  \bibinfo{author}{\bibfnamefont{P.}~\bibnamefont{Gori-Giorgi}},
  \bibinfo{journal}{Phys. Rev. B} \textbf{\bibinfo{volume}{{87}}},
  \bibinfo{pages}{115146} (\bibinfo{year}{2013}).

\bibitem[{\citenamefont{Mendl et~al.}(2014)\citenamefont{Mendl, Malet, and
  Gori-Giorgi}}]{MenMalGor-PRB-14}
\bibinfo{author}{\bibfnamefont{C.~B.} \bibnamefont{Mendl}},
  \bibinfo{author}{\bibfnamefont{F.}~\bibnamefont{Malet}}, \bibnamefont{and}
  \bibinfo{author}{\bibfnamefont{P.}~\bibnamefont{Gori-Giorgi}},
  \bibinfo{journal}{Phys. Rev. B} \textbf{\bibinfo{volume}{89}},
  \bibinfo{pages}{125106} (\bibinfo{year}{2014}).

\bibitem[{\citenamefont{Lieb}(1983)}]{Lie-IJQC-83}
\bibinfo{author}{\bibfnamefont{E.~H.} \bibnamefont{Lieb}},
  \bibinfo{journal}{Int. J. Quantum. Chem.} \textbf{\bibinfo{volume}{{24}}},
  \bibinfo{pages}{24} (\bibinfo{year}{1983}).

\bibitem[{\citenamefont{Seidl}(1999)}]{Sei-PRA-99}
\bibinfo{author}{\bibfnamefont{M.}~\bibnamefont{Seidl}},
  \bibinfo{journal}{Phys. Rev. A} \textbf{\bibinfo{volume}{{60}}},
  \bibinfo{pages}{4387} (\bibinfo{year}{1999}).

\bibitem[{\citenamefont{Seidl et~al.}(2007)\citenamefont{Seidl, Gori-Giorgi,
  and Savin}}]{SeiGorSav-PRA-07}
\bibinfo{author}{\bibfnamefont{M.}~\bibnamefont{Seidl}},
  \bibinfo{author}{\bibfnamefont{P.}~\bibnamefont{Gori-Giorgi}},
  \bibnamefont{and} \bibinfo{author}{\bibfnamefont{A.}~\bibnamefont{Savin}},
  \bibinfo{journal}{Phys. Rev. A} \textbf{\bibinfo{volume}{{75}}},
  \bibinfo{pages}{042511} (\bibinfo{year}{2007}).

\bibitem[{\citenamefont{Buttazzo et~al.}(2012)\citenamefont{Buttazzo, {De
  Pascale}, and Gori-Giorgi}}]{ButDepGor-PRA-12}
\bibinfo{author}{\bibfnamefont{G.}~\bibnamefont{Buttazzo}},
  \bibinfo{author}{\bibfnamefont{L.}~\bibnamefont{{De Pascale}}},
  \bibnamefont{and}
  \bibinfo{author}{\bibfnamefont{P.}~\bibnamefont{Gori-Giorgi}},
  \bibinfo{journal}{Phys. Rev. A} \textbf{\bibinfo{volume}{{85}}},
  \bibinfo{pages}{062502} (\bibinfo{year}{2012}).

\bibitem[{\citenamefont{Cotar et~al.}(2013)\citenamefont{Cotar, Friesecke, and
  Kl\"uppelberg}}]{CotFriKlu-CPAM-13}
\bibinfo{author}{\bibfnamefont{C.}~\bibnamefont{Cotar}},
  \bibinfo{author}{\bibfnamefont{G.}~\bibnamefont{Friesecke}},
  \bibnamefont{and}
  \bibinfo{author}{\bibfnamefont{C.}~\bibnamefont{Kl\"uppelberg}},
  \bibinfo{journal}{Comm. Pure Appl. Math.} \textbf{\bibinfo{volume}{66}},
  \bibinfo{pages}{548} (\bibinfo{year}{2013}).

\bibitem[{\citenamefont{Cotar et~al.}(2014)\citenamefont{Cotar, Friesecke, and
  Pass}}]{CodFriPas-INC-14}
\bibinfo{author}{\bibfnamefont{C.}~\bibnamefont{Cotar}},
  \bibinfo{author}{\bibfnamefont{G.}~\bibnamefont{Friesecke}},
  \bibnamefont{and} \bibinfo{author}{\bibfnamefont{B.}~\bibnamefont{Pass}}, in
  \emph{\bibinfo{booktitle}{Calculus of Variations and Partial Differential
  Equations}} (\bibinfo{publisher}{Springer}, \bibinfo{address}{Berlin
  Heidelberg}, \bibinfo{year}{2014}), pp. \bibinfo{pages}{1--26}.

\bibitem[{\citenamefont{Friesecke et~al.}(2013)\citenamefont{Friesecke, Mendl,
  Pass, Cotar, and Kl\"uppelberg}}]{FriMenPasCotKlu-JCP-13}
\bibinfo{author}{\bibfnamefont{G.}~\bibnamefont{Friesecke}},
  \bibinfo{author}{\bibfnamefont{C.~B.} \bibnamefont{Mendl}},
  \bibinfo{author}{\bibfnamefont{B.}~\bibnamefont{Pass}},
  \bibinfo{author}{\bibfnamefont{C.}~\bibnamefont{Cotar}}, \bibnamefont{and}
  \bibinfo{author}{\bibfnamefont{C.}~\bibnamefont{Kl\"uppelberg}},
  \bibinfo{journal}{J. Chem. Phys.} \textbf{\bibinfo{volume}{139}},
  \bibinfo{pages}{164109} (\bibinfo{year}{2013}).

\bibitem[{\citenamefont{Colombo et~al.}(2014)\citenamefont{Colombo, {De
  Pascale}, and {Di Marino}}}]{ColDepDiM-CJM-14}
\bibinfo{author}{\bibfnamefont{M.}~\bibnamefont{Colombo}},
  \bibinfo{author}{\bibfnamefont{L.}~\bibnamefont{{De Pascale}}},
  \bibnamefont{and} \bibinfo{author}{\bibfnamefont{S.}~\bibnamefont{{Di
  Marino}}}, \bibinfo{journal}{Can. J. Math.} \textbf{\bibinfo{volume}{{}}},
  \bibinfo{pages}{accepted} (\bibinfo{year}{2014}).

\bibitem[{\citenamefont{Pass}(2013{\natexlab{a}})}]{Pas-NL-13}
\bibinfo{author}{\bibfnamefont{B.}~\bibnamefont{Pass}},
  \bibinfo{journal}{Nonlinearity} \textbf{\bibinfo{volume}{{26}}},
  \bibinfo{pages}{2731} (\bibinfo{year}{2013}{\natexlab{a}}).

\bibitem[{\citenamefont{Pass}(2013{\natexlab{b}})}]{Pas-JFA-13}
\bibinfo{author}{\bibfnamefont{B.}~\bibnamefont{Pass}}, \bibinfo{journal}{J.
  Func. Analysis} \textbf{\bibinfo{volume}{{264}}}, \bibinfo{pages}{947}
  (\bibinfo{year}{2013}{\natexlab{b}}).

\bibitem[{\citenamefont{Ghoussoub and Moameni}(2014)}]{GhoMoa-GFA-14}
\bibinfo{author}{\bibfnamefont{N.}~\bibnamefont{Ghoussoub}} \bibnamefont{and}
  \bibinfo{author}{\bibfnamefont{A.}~\bibnamefont{Moameni}},
  \bibinfo{journal}{Geom. Funct. Anal.} \textbf{\bibinfo{volume}{{24}}},
  \bibinfo{pages}{1129} (\bibinfo{year}{2014}).

\bibitem[{\citenamefont{Kim and Pass}(2014)}]{KimPas-SJMA-14}
\bibinfo{author}{\bibfnamefont{Y.-H.} \bibnamefont{Kim}} \bibnamefont{and}
  \bibinfo{author}{\bibfnamefont{B.}~\bibnamefont{Pass}},
  \bibinfo{journal}{SIAM J. Math. Anal.} \textbf{\bibinfo{volume}{{46}}},
  \bibinfo{pages}{1538} (\bibinfo{year}{2014}).

\bibitem[{\citenamefont{Colombo and {Di Marino}}(2013)}]{ColDiM-INC-13}
\bibinfo{author}{\bibfnamefont{M.}~\bibnamefont{Colombo}} \bibnamefont{and}
  \bibinfo{author}{\bibfnamefont{S.}~\bibnamefont{{Di Marino}}}, in
  \emph{\bibinfo{booktitle}{Annali di Matematica Pura ad Applicata}}
  (\bibinfo{publisher}{Springer}, \bibinfo{address}{Berlin Heidelberg},
  \bibinfo{year}{2013}), pp. \bibinfo{pages}{1--14}.

\bibitem[{\citenamefont{Mirtschink et~al.}(2012)\citenamefont{Mirtschink,
  Seidl, and Gori-Giorgi}}]{MirSeiGor-JCTC-12}
\bibinfo{author}{\bibfnamefont{A.}~\bibnamefont{Mirtschink}},
  \bibinfo{author}{\bibfnamefont{M.}~\bibnamefont{Seidl}}, \bibnamefont{and}
  \bibinfo{author}{\bibfnamefont{P.}~\bibnamefont{Gori-Giorgi}},
  \bibinfo{journal}{J. Chem. Theory Comput.} \textbf{\bibinfo{volume}{8}},
  \bibinfo{pages}{3097} (\bibinfo{year}{2012}).

\bibitem[{\citenamefont{R\"as\"anen et~al.}(2011)\citenamefont{R\"as\"anen,
  Seidl, and Gori-Giorgi}}]{RasSeiGor-PRB-11}
\bibinfo{author}{\bibfnamefont{E.}~\bibnamefont{R\"as\"anen}},
  \bibinfo{author}{\bibfnamefont{M.}~\bibnamefont{Seidl}}, \bibnamefont{and}
  \bibinfo{author}{\bibfnamefont{P.}~\bibnamefont{Gori-Giorgi}},
  \bibinfo{journal}{Phys. Rev. B} \textbf{\bibinfo{volume}{{83}}},
  \bibinfo{pages}{195111} (\bibinfo{year}{2011}).

\bibitem[{\citenamefont{Cremon et~al.}(2010)\citenamefont{Cremon, Bruun, and
  Reimann}}]{CreBruRei-PRL-10}
\bibinfo{author}{\bibfnamefont{J.~C.} \bibnamefont{Cremon}},
  \bibinfo{author}{\bibfnamefont{G.~M.} \bibnamefont{Bruun}}, \bibnamefont{and}
  \bibinfo{author}{\bibfnamefont{S.~M.} \bibnamefont{Reimann}},
  \bibinfo{journal}{Phys. Rev. Lett.} \textbf{\bibinfo{volume}{{105}}},
  \bibinfo{pages}{255301} (\bibinfo{year}{2010}).

\bibitem[{\citenamefont{Mendl and Lin}(2013)}]{MenLin-PRB-13}
\bibinfo{author}{\bibfnamefont{C.~B.} \bibnamefont{Mendl}} \bibnamefont{and}
  \bibinfo{author}{\bibfnamefont{L.}~\bibnamefont{Lin}},
  \bibinfo{journal}{Phys. Rev. B} \textbf{\bibinfo{volume}{87}},
  \bibinfo{pages}{125106} (\bibinfo{year}{2013}).

\bibitem[{\citenamefont{Chen et~al.}(2014)\citenamefont{Chen, Friesecke, and
  Mendl}}]{CheFriMen-JCTC-14}
\bibinfo{author}{\bibfnamefont{H.}~\bibnamefont{Chen}},
  \bibinfo{author}{\bibfnamefont{G.}~\bibnamefont{Friesecke}},
  \bibnamefont{and} \bibinfo{author}{\bibfnamefont{C.~B.} \bibnamefont{Mendl}},
  \bibinfo{journal}{J. Chem. Theory Comput.} \textbf{\bibinfo{volume}{10}},
  \bibinfo{pages}{4360} (\bibinfo{year}{2014}).

\bibitem[{\citenamefont{Gori-Giorgi
  et~al.}(2009{\natexlab{b}})\citenamefont{Gori-Giorgi, Vignale, and
  Seidl}}]{GorVigSei-JCTC-09}
\bibinfo{author}{\bibfnamefont{P.}~\bibnamefont{Gori-Giorgi}},
  \bibinfo{author}{\bibfnamefont{G.}~\bibnamefont{Vignale}}, \bibnamefont{and}
  \bibinfo{author}{\bibfnamefont{M.}~\bibnamefont{Seidl}}, \bibinfo{journal}{J.
  Chem. Theory Comput.} \textbf{\bibinfo{volume}{{5}}}, \bibinfo{pages}{743}
  (\bibinfo{year}{2009}{\natexlab{b}}).

\bibitem[{\citenamefont{Fang and Englert}(2011)}]{FanEng-PRA-11}
\bibinfo{author}{\bibfnamefont{B.}~\bibnamefont{Fang}} \bibnamefont{and}
  \bibinfo{author}{\bibfnamefont{B.-G.} \bibnamefont{Englert}},
  \bibinfo{journal}{Phys. Rev. A} \textbf{\bibinfo{volume}{{83}}},
  \bibinfo{pages}{052517} (\bibinfo{year}{2011}).

\bibitem[{\citenamefont{Lahrz et~al.}(2014)\citenamefont{Lahrz, Lemeshko, and
  Mathey}}]{LahMikMat-arx-14}
\bibinfo{author}{\bibfnamefont{M.}~\bibnamefont{Lahrz}},
  \bibinfo{author}{\bibfnamefont{M.}~\bibnamefont{Lemeshko}}, \bibnamefont{and}
  \bibinfo{author}{\bibfnamefont{L.}~\bibnamefont{Mathey}},
  \bibinfo{journal}{arXiv:1412.8472S} \textbf{\bibinfo{volume}{{}}}
  (\bibinfo{year}{2014}).

\bibitem[{\citenamefont{Bhongale et~al.}(2013)\citenamefont{Bhongale, Mathey,
  Zhao, Yelin, and Lemeshko}}]{BhoMatZhaYelLem-PRL-13}
\bibinfo{author}{\bibfnamefont{S.~G.} \bibnamefont{Bhongale}},
  \bibinfo{author}{\bibfnamefont{L.}~\bibnamefont{Mathey}},
  \bibinfo{author}{\bibfnamefont{E.}~\bibnamefont{Zhao}},
  \bibinfo{author}{\bibfnamefont{S.}~\bibnamefont{Yelin}}, \bibnamefont{and}
  \bibinfo{author}{\bibfnamefont{M.}~\bibnamefont{Lemeshko}},
  \bibinfo{journal}{Phys. Rev. Lett.} \textbf{\bibinfo{volume}{{110}}},
  \bibinfo{pages}{155301} (\bibinfo{year}{2013}).

\bibitem[{\citenamefont{Huang et~al.}(2014)\citenamefont{Huang, Lahrz, and
  Mathey}}]{HuaLahMat-arx-13}
\bibinfo{author}{\bibfnamefont{W.-M.} \bibnamefont{Huang}},
  \bibinfo{author}{\bibfnamefont{M.}~\bibnamefont{Lahrz}}, \bibnamefont{and}
  \bibinfo{author}{\bibfnamefont{L.}~\bibnamefont{Mathey}},
  \bibinfo{journal}{arXiv:1311.1947v2} \textbf{\bibinfo{volume}{{}}}
  (\bibinfo{year}{2014}).

\end{thebibliography}

\end{document}